\documentclass[twocolumn,nofootinbib,superscriptaddress,reprint]{revtex4-1}

\pdfoutput=1

\usepackage{slashed}

\usepackage[normalem]{ulem}

\usepackage{color}

\usepackage{epsfig}
\usepackage{epstopdf}
\usepackage{epsf}
\input{epsf.sty}
\usepackage{graphicx,amsmath,amssymb}
\usepackage{epsfig}
\allowdisplaybreaks

\def\ei{\end{itemize}}
\def\be{\begin{equation}}
\def\ee{\end{equation}}
\newcommand{\bea}{\begin{eqnarray}}
\newcommand{\eea}{\end{eqnarray}}

\usepackage{bbm,bm,amsmath,amssymb}

\usepackage{hyperref}

\begin{document}

\title{Chiral Gravitational Waves and Baryon Superfluid Dark Matter} 

\author{Stephon Alexander}
\email{stephon\_alexander@brown.edu}
\author{ Evan McDonough}
\email{evan\_mcdonough@brown.edu}
\affiliation{Department of Physics, Brown University,\\ 
Barus and Holley, Providence, RI, USA. 02903.\\}
\author{David N. Spergel}
\email{dns@astro.princeton.edu}
\affiliation{Department of Astrophysical Sciences, Princeton University, Princeton, NJ 08544, USA}
\affiliation{Center for Computational Astrophysics, Flatiron Institute, New York, NY 10003, USA}

 \begin{abstract}
We develop a unified model of darkgenesis and baryogenesis involving strongly interacting dark quarks, utilizing the gravitational anomaly of chiral gauge theories. In these models, both the visible and dark baryon asymmetries are generated by the gravitational anomaly induced by the presence of chiral primordial gravitational waves.  We provide a concrete model of an SU(2) gauge theory with two massless quarks. In this model, the dark quarks condense and form a dark baryon charge superfluid (DBS), in which the Higgs-mode acts as cold dark matter. We elucidate the essential features of this dark matter scenario and discuss its phenomenological prospects.
\end{abstract}

\maketitle

\section{Introduction}

Despite ample evidence for dark matter, the precise particle nature of dark matter remains elusive. For the last three decades, the guiding principle for much of dark matter research has been the so-called \emph{WIMP miracle}. However, as WIMPS continue to evade detection, it is natural to consider alternatives to the WIMP paradigm.

One possibility is to seek guidance from \emph{cosmological coincidence} in the energy density budget of the universe, namely that $\frac{\Omega_{DM}}{\Omega_b} \simeq 5$. The paradigm of composite dark matter \cite{Kribs:2016cew} posits that dark matter may be strongly interacting, and predominantly confined into \emph{dark baryons}, while the paradigm of asymmetric dark matter \cite{Petraki:2013wwa} uses the above relation to relate the dark matter mass and the dark particle asymmetry to those in the visible sector,  $n_{b}m_b \simeq n_{d}m_{d}$.  In this work we will directly tie the genesis of baryon number to dark matter number density, while the dark matter mass will be determined by the details of the microscopic theory.

To do this, we put forward a new approach utilizing the gravitational anomaly of gauge theories containing chiral fermions \cite{AlvarezGaume:1983ig}:
\be
\label{eq:anomaly}
\partial_\mu \left( \sqrt{-g}  j^\mu _{L,R}\right) =\pm  \frac{  N_{L,R} }{12} \frac{1}{16 \pi^2} R \tilde{R} .
\ee
where $N_{L,R}$ are number of left,right-handed fermion species, the $+/-$ corresponds to $L/R$, and $R \tilde{R}$ is a contraction of two Reimann tensors with a Levi-Civita tensor. This anomaly was a key ingredient for an inflationary leptogenesis model in \cite{Alexander:2004us} and has subsequently been studied in many works e.g. \cite{lepto,adshead}. In our work we extend this mechanism to both the visible and dark sectors, leading to simultaneous leptogenesis in the visible sector and \emph{baryogenesis} in the dark sector. 

This applicability of this is independent of many details of the dark matter model. The requirements are that (1) there is a early universe production of chiral gravitational waves, and (2) the dark matter model be chiral, i.e. that the number of left-handed fields and right-handed fields differ, $N_L - N_R \neq 0$.   For the standard model, the requisite chirality $N_{L} \neq N_R$ arises due to the left-handed neutrinos\footnote{In fact, the resulting baryon asymmetry is unchanged in the presence of a mass for neutrinos \cite{Alexander:2004us}\cite{adshead}.}. Thus a straightforward realization of unified dark and visible genesis is a dark standard model, occurring along the same lines as in the original formulation of leptogenesis from gravity waves \cite{Alexander:2004us}. 

Furthermore, this mechanism does not require any direct interactions with the standard model. With this in mind, we present a concrete working model:  SU(2) gauge theory with two massless quarks. Primordial chiral gravitational waves generate a large particle-antiparticle asymmetry for the dark quarks, inducing a chemical potential. Subsequently, similar to QCD at low-temperature and high density (see e.g. \cite{Kogut:2004su,Alford:2007xm}), the quarks condense and form a \emph{dark baryon charge superfluid} (DBS)\footnote{ We note the relation of this work to some existing works: ``Cogenesis'' of dark and visible matter has been considered in \cite{Kamada:2012ht}, chiral fermions have played a large role in the Darkogenesis scenario \cite{Shelton:2010ta}, and the Hypercharge chiral anomaly of the Standard Model, also used to generate a visible asymmetry in e.g. \cite{lepto2} and references therein, has been used to engineer a dark asymmetry \cite{Cado:2016kdp}. In contrast to those works, our mechanism requires no messenger sector to transfer the asymmetry between the standard model and the dark sector. Chiral anomalies during inflation were incorporated in \cite{Barrie:2015axa}, but in contrast to that work, the dark matter here is not required to share a common global charge with the Standard Model.}
. 

The low energy degrees of freedom of the condensate are a Higgs-mode and a Goldstone boson. The Higgs-mode serves as a cold dark matter candidate, with a mass that is bounded from above by a function of the reheating temperature of the universe. This allows dark matter in the range of so called ultra-light axions \cite{Marsh:2015xka}, though heavier, e.g. GeV, masses are possible. This opens up a rich phenomenology distinct from that of axions\cite{Marsh:2015xka}, Bose-Einstein Condensates  \cite{Sikivie:2009qn}, and superfluid dark matter \cite{Berezhiani:2015bqa}. We highlight these differences in Section \ref{SEC.Discussion}.

\section{Non-Abelian Dark Matter Genesis}
\label{sec:genesis}

\subsection{Overview of the Mechanism}
\label{sec:overview}

The dark matter scenario under consideration begins with CP-violation during inflation, which sources birefringent gravitational waves.  This naturally leads to a dark matter particle-antiparticle asymmetry via. the gravitational anomaly \eqref{eq:anomaly}. A prototypical example of a model where this can be realized is a non-abelian gauge theory with a net number of chiral fermions, e.g. a dark standard model. 

It was realized in  \cite{Alexander:2004us} that chiral gravitational waves can lead to leptogenesis via the anomaly \eqref{eq:anomaly}\footnote{Further work connecting chiral gravitational waves to observables can be found in \cite{AG,AM,AY,AF}.}. The Sakharov conditions: (i) B-conservation violation, (ii) CP violation, and (iii) non-equilibrium, are simultaneously broken by the non-trivial topology of the gravitational field, as measured by $\int R \tilde{R}$. This leads to a violation of lepton number conservation, which is subsequently transferred to baryon number violation via sphaleron processes.

A key detail for the present work is the application of the anomaly equation \eqref{eq:anomaly} to a theory of multiple decoupled gauge theories. In each sector, labelled by $i$, baryon number ($B_i$) conservation arises as an accidental global symmetry $U(1)_{B_i}$. This is broken by the axial-gravitational anomaly, generated by a triangle diagram of the axial current $j_{(i)\mu A}$ and two insertions of the gravitational stress tensor. This leads to a set of independent anomaly equations,  
\be
\partial_\mu \left( \sqrt{-g}  j^{(i)\mu} _{B}\right) = \frac{\left( N_L ^{(i)} - N_R ^{(i)} \right)}{12} \frac{1}{16 \pi^2} R \tilde{R}  ,
\ee
where $j^{(i)\mu} _{B}$ is the baryon number current in the $i^{th}$ gauge sector.  This allows for simultaneous violation of standard model lepton number conservation and dark sector `dark baryon' or `dark lepton' number conservation, and thus a generation of standard model and dark sector asymmetries.

Independent of the precise details of the model, this leads to a dark matter asymmetry proportional to that in the standard model,
\be
\label{eq:etaD}
\eta_{D} =    \left(  \frac{79}{28} \frac{N_L ^{D} - N_{R} ^{D}}{3} \right) \eta_b
\ee
where $\eta_{b,D}$ are the baryon asymmetries in the standard model and dark sectors respectively, $N_{L,R}^{D}$ refers to the number left or right fermions in the dark sector, and the factor of $79/28$ accounts for the transfer of lepton number to baryon number via sphaleron processes in the Standard Model \cite{sphaleron}.

The gravitational waves responsible for this asymmetry can be measured in the B-mode polarization of the CMB. Demanding $\eta_b/s$ takes on its observed value then leads a lower-bound on the tensor-to-scalar ratio $r$. The bound is model dependent, and in the example of \cite{Caldwell:2017chz} is given by 
\be
r \gtrsim 10^{-2} ,
\ee
well within the range of detection by next-generation CMB experiments \cite{Abazajian:2016yjj}.

We can interpret this as follows: If $r$ is observed to be less then this value, then \cite{Caldwell:2017chz} requires an additional mechanism to produce $\eta_b$, and its use in our scenario may require an additional mechanism to produce a sufficient amount of dark matter.

\subsection{Dark SU(2)}
\label{sec:DarkSU2}

The reliance of darkgenesis on the gravitational anomaly places strict constraints on the gauge sector and subsequent state that the dark matter evolves into in the late universe.  The gravitational anomaly produces dark quarks only if there is an imbalance between left and right handed fermions, $N_{L}\neq N_{R}$.  This will naturally introduce gauge anomalies which will need to be cancelled \cite{Harvey:2005it}.  For $SU(N)$ gauge theories with $N>2$, this leads to an $SU(N)^3$ anomaly, thus uniquely picking out SU(2) as the gauge group. This SU(2) gauge theory is constrained by the Witten anomaly \cite{Witten:1982fp} to have an even number of doublets of Weyl fermions, and hence the minimum number of flavours of quarks in this model is two. 

There are many possible sources for the requisite chiral gravitational waves: a direct coupling of the inflaton to gravity \cite{Lue:1998mq, Alexander:2004us}, Abelian gauge fields coupled to the inflaton \cite{Sorbo:2011rz} or a spectator field \cite{Ferreira:2014zia,Ferreira:2015omg,Namba:2015gja}, and non-Abelian gauge fields with \cite{Adshead:2013qp, Adshead:2013nka, Maleknejad:2016qjz,Maleknejad:2014wsa, Maleknejad:2016dci,Caldwell:2017chz, GWsSU2} or without  \cite{Bielefeld:2015daa} a direct coupling to the inflaton or spectator field. In all of these cases, the slow-roll inflation consistency relation $n_t = -r/8$ and the Lyth bound \cite{Lyth:1996im} are violated. 

For concreteness we will focus on the production of gravitational waves in the chromo-natural inflation \cite{Noorbala:2012fh,Adshead:2012kp} scenario studied in \cite{Caldwell:2017chz}, wherein a non-abelian gauge field (here taken to be $SU(2)$ for simplicity) is coupled to the inflaton via an axionic coupling. We add to this the dark sector, consisting of a dark $SU(2)$ and two left-handed Weyl fermions.

The action for this system is given by:
\bea
S =&& \int d^4 x \; \sqrt{-g} \left[ \right. \frac{M_{Pl^2}}{2}R - \frac{1}{2}(\partial \phi)^2 - V(\phi) - \frac{1}{4} F^2   - \frac{\phi}{M} F \tilde{F} \nonumber \\
 && - \psi^I \slashed{D} \psi_I   - \frac{1}{4} G^2 + \mathcal{L}_{SM} \left. \right] ,
\eea
where the first line describes inflation: $\phi$ is the inflaton, $F^a_{\mu\nu} = \partial_{[\mu} A^a _{\nu ]} - g_A \epsilon^{abc} A_{b\mu} A_{c\nu}$ is the non-Abelian field strength with coupling $g_A$, and $\tilde{F}$ is the Hodge dual of $F$. The dark sector is given by the first two terms of the second line, with $I=1,2$ is the flavour index of the fermions $\psi_I$, and the dark SU(2) field strength is denoted as  $G^a_{\mu\nu} = \partial_{[\mu} B^a _{\nu ]} - g \epsilon^{abc} B_{b\mu} B_{c\nu}$. The third term on the second line is Lagrangian of the Standard Model.

The equations of motion describing this system are given by the Friedman equation, the inflaton equation of motion, the gauge field equation of motion, the Dirac equation for the fermions, and the chiral anomaly, 
\bea
&& \frac{H^2}{3 M_{Pl}^2} = \sum_i \rho_i \;\; ,\;\;\ddot{\phi} + 3 H \dot{\phi} + V_{, \phi} = \frac{1}{M} F \tilde{F} \\
&& \label{eq:gaugefields}\partial_\mu (\sqrt{-g} F^{a\mu \nu}) + \frac{1}{M} \partial_{\mu} (\sqrt{-g} \phi \tilde{F}^{a \mu \nu}) =0 \\
&& \partial_\mu (\sqrt{-g} G^{a\mu \nu})=0, \\
&& \label{eq:fermions} \slashed{D}_\mu \psi_I =0 \;,\; \partial_\mu \left( \sqrt{-g}  j^\mu _{5}\right) = \frac{1}{192 \pi^2} R \tilde{R} +  \frac{1}{32 \pi^2}G \tilde{G} . 
\eea
In addition, there is the equation of motion for gravitational waves:
\be
h_{ij} ' + 2 \mathcal{H} h_{ij}' - \nabla^2 h_{ij} = 2 a^2 \Pi_{IJ}  ,
\ee
where $\Pi_{IJ}$ is the (transverse-traceless) anistropic stress sourced by the inflaton, gauge field, and fermions.  There is also an equation of motion for the scalar metric fluctuations, which will we will not study here.

The inflationary dynamics in the scenario \cite{Caldwell:2017chz} occurs via a balancing of the inflaton potential with the energy density of the gauge field $A$ in the isotropic configuration,
\be
A_i ^a = \sigma(t) \delta^a _i .
\ee
This leads to a non-vanishing $F^2$ and $F \tilde{F}$ at the background level, and inflation occurs in the effective potential $V_{eff} = V(\phi) - \phi F \tilde{F}/M$.

The source term for gravitational waves will receive contributions from the inflaton, fermion, and gauge field fluctuations. The inflaton sources an anisotropic stress at second-order in perturbation theory, but absent any amplification of gradients (e.g. via preheating \cite{Dufaux:2007pt}), this is vastly subdominant to the production due to gauge fields. As for the fermions, while their production will source gravitational waves \cite{Anber:2016yqr}, this effect is also subdominant: In \cite{Anber:2016yqr}, which utilizes a direct inflaton-fermion coupling, a large number of fermion species is required to have any appreciable effect on the gravitational wave spectrum. Thus the dominant source of gravitational wave production will be from the tensorial gauge field fluctuations studied in  \cite{Caldwell:2017chz, Maleknejad:2014wsa}, and the mechanism of  \cite{Caldwell:2017chz, Maleknejad:2014wsa} can proceed unimpeded. 

The subtle piece of new physics introduced in this model is the dynamics of the fermions during and after inflation. This is dictated by equation \eqref{eq:fermions}, from which we will be able to compute the number density of dark quarks and asymmetry.  The axial anomaly has contributions from both $G\tilde{G}$ and $R \tilde{R}$ \cite{Harvey:2005it}, however since $G \tilde{G}$ is not directly produced by inflation, the dominant source will be $R \tilde{R}$. The resulting asymmetry is given by equation \eqref{eq:etaD}.

We review the calculation of the gravitational wave spectrum and the resulting Standard Model leptogenesis in Appendix \ref{app:GWs}. The result can be expressed as a ratio with the entropy density of the universe, $s = 2 g_{*s}\pi^2 T^3/45$, as
\be
\frac{|\langle n_{\ell} \rangle|}{s} = 2.45 \times 10^{-10} .
\ee
where the parameters chosen are $g_{*s}= 1.4 \times 10^{-3}$, $M=1.7 \times 10^{-4} M_{Pl}$, and the inflation model is given by and the fiducial inflation model $V =m^{4}(\phi/m)^n /n$ with $m = 1.7 \times 10^{-3}M_{Pl}$ and $n=1/8$.

The corresponding quantity in the dark sector is simply
\be
\label{nDs}
\frac{|\langle n_{D} \rangle |}{s} = \frac{2}{3} \frac{|\langle n_{\ell} \rangle|}{s} = 1.63 \times 10^{-10} ,
\ee
and hence a sizeable particle-antiparticle asymmetry is generated in the dark sector.

\section{Dark Baryon Charge Superfluid (DBS) Dark Matter}
\label{sec:DBS}

The previous section established an initial condition for the cosmological evolution of the dark sector. We now proceed to develop this model in detail.

The post-inflationary state of the dark sector after inflation is a non-thermal population of quarks, at finite density and hence chemical potential. This naturally leads to the formation of a ``color-flavor locked'' condensate, which is, in this case, a \emph{superfluid}. This occurs due to single-gluon exchange between the dark quarks, which analogous to the phonon-mediated interaction of BCS theory, provides an attractive interaction between quarks. For a review of color superconductivity, we refer the reader to \cite{Alford:2007xm}. 

Before proceeding, we first perform some basic consistency checks. Since the condensate is only stable at low temperatures, it is of the utmost importance that the condensate forms before the system can thermalize. The dominant interactions between the quarks are attractive and repulsive single-gluon exchange diagrams, the first of which leads to the pairing of quarks and the formation of the condensate, while the second of which will tend to thermalize the quarks (as for example electrons undergoing Compton scattering in the early universe).  

The ratio of the attractive to repulsive scattering amplitudes is determined by the number of colors, and is given by $\mathcal{M}_a / \mathcal{M}_r = (N_c + 1)/(N_c - 1)$ (see e.g. \cite{Ruester:2006yh}). The interaction rates of the two processes, $\Gamma = n \sigma v$, then gives the ratio of time-scales (for $N_c =2$) $t_a / t_r = 1/9 $, indicating that the time scale for the formation of the condensate is roughly ten times shorter then the time scale for repulsive scatterings. Since many successive scatterings are needed to thermalize the quark population, we conclude that condensate will form well before the system can thermalize.

Moreover, in order for the formation to occur in the early universe, the interaction rate of gluon exchange must be greater then the Hubble expansion rate $H$. Similar to the canonical calculation for determining thermal equilibrium (see e.g. \cite{Mukhanov:2005sc}), the ratio scales as $\Gamma/H \propto a(t)$. More precisely, for gluon exchange at the Fermi momentum $k_F$,
 \be
\frac{\Gamma}{H} = \frac{n_e \alpha^2}{k_{F(e)}^2 H_e} \left( \frac{a(t)}{a(t_e)} \right).
\ee
where the subscript $e$ denotes the value at the end of inflation, and $\alpha = g ^2/4\pi$ is the coupling. The number density $n_e$ can be estimated by assuming most fluctuations have $k = k_F$, such that the energy density in DM is given by $\rho_{DM} = n k_F$. An upper bound on $k_{F}$ is $H_{e}$, since the production stops at the end of inflation. Moreover, $H_e$ is the total energy density of the universe after inflation, $H_e = \sqrt{\rho_e} / \sqrt{3}M_{Pl} $. Therefore we have
\be
\frac{\Gamma}{H} = \frac{\rho_{DM(e)}}{\rho_e} \frac{M_{Pl} ^2}{H_e ^2} \alpha^2 \left( \frac{a(t)}{a(t_e)} \right) .
\ee
The first ratio on the LHS is roughly $10^{-5}$ and the second ratio is $ 10^{20}$ if $H_e =  10^{-10} M_{Pl}$, and this gives
\be
\frac{\Gamma}{H}|_{t_e} = 10^{15} \alpha^2 \gg 1
\ee
Thus the DBS can form any time after inflation, provided the coupling is not too small.

The final cosmological issue to address before proceeding is the correlation length of the condensate. This situation here is markedly different from axion condensate models \cite{Sikivie:2009qn}, where a super-Hubble correlation length is required at all times in order for the particle excitations to be ignored on cosmological scales. In contrast to this, the formation of the diquark condensate is indicated by a nonzero VEV of a gap $\Delta$, which can occur independently in separated spatial regions.  As there is only one vacuum solution for $\Delta$, its value will be uniform throughout the universe. Hence there is no constraint on the model from the correlation length. 

\subsection{The superfluid phase of massless SU(2) QCD}

\label{sec:SU2DBS}

We now take a quick tour of the phase diagram of gauge theories. The formation of a color superconductor (or in the SU(2) case, a superfluid), occurs when the chemical potential $\mu$ is greater then the mass of the lightest baryon $m_B$. This leads to a gap $\Delta$, and provided that the temperature of the system is below the gap, a stable condensate is formed. Above the critical temperature $T_c \simeq \Delta$ \footnote{Quantitatively, in QCD $T_c = 0.57 \Delta  $ \cite{Alford:2007xm}.}, the theory deconfines into a plasma, and below the critical chemical potential $\mu_c \simeq m_{B}$ the system is in a hadronic phase (or else plasma if the temperature is large enough).

The case of massless chiral SU(2) is special, as the lightest baryon is in fact \emph{massless}. This can be understood from the fact that chiral symmetry is unbroken, along with the degeneracy between hadron and meson masses in SU(2) \cite{Kogut:2004su}. The first fact implies that the lightest baryon has a mass equal to the pion, while the second fact implies that the pion is massless. Hence the massless SU(2) theory is unstable to the formation of a condensate \emph{at any non-zero value of the chemical potential}.

The Lagrangian describing this system is given by
\be \mathcal{L}  =  \bar{\psi}(i\gamma^{\mu} D_{\mu} - \mu_\mu \gamma^{\mu}) \psi - \frac{1}{4} G^{\mu\nu}_{a} {G^{\mu\nu}}^{a}  \ee
where $\mu^\mu$ is the chemical potential four-vector, which we take to be $\mu^0=\mu$ and $\mu^i=0$ in the rest frame of the CMB. A four-fermion interaction $\cal{L}\rm_{int}$  arises after integrating out the dark gluons \cite{Alford:1998mk,Alford:1997zt, Alford:2007xm}, with the standard QCD interaction vertex $\bar{\psi}_{i  \alpha} \gamma^{\mu } A_{\mu} ^a T^{a \alpha \beta}  \psi_{j   \beta} \delta^{ij}$, and using the SU(N) identity $2 T^a _{\alpha \beta} T^a _{\gamma \delta} = \delta_{\alpha \delta} \delta_{\beta \gamma} - (1/N) \delta_{\alpha \beta} \delta_{\gamma \delta} $,  
\be
 \label{eq:int} 
    \mathcal{L}_{int}  = g_{4f}\,  \bar{\psi}_{i \alpha } \gamma^\mu \psi_{j \beta} \bar{\psi}_{k \gamma} \gamma_\mu \psi_{l \delta} \delta^{ij} \delta^{kl}( 2 \delta^{\alpha \delta}  \delta^{\beta \gamma} - \delta^{\alpha \beta} \delta^{\gamma \delta}  ) 
 \ee
where $i,j,k,l$ are flavor indices, $\alpha, \beta, \gamma, \delta$ are color indices, and the spinor indices are suppressed. This interaction is attractive for $qq\rightarrow qq$ scattering when the color state is antisymmetric (e.g. $(\alpha \beta - \beta \alpha)/\sqrt{2})$, which leads to the pairing of quarks.

The formation of the condensate is encoded in a non-zero vacuum expectation value of a diquark state, which is antisymmetric in both color and flavor (the antisymmetry in flavor following from Fermi-Dirac statistics). Writing the fermions as two-component Weyl spinors, the condensate is given by\cite{Schafer:2002ms}
\be
\label{eq:psiLpsiL} 
\langle {\psi^{ i}_{L \alpha a}\psi^{j}_{L \beta c}\epsilon^{a c}} \rangle =  \Delta^{ij}_{\alpha\beta} = \Delta \epsilon^{ij} \epsilon_{\alpha \beta} = \Delta(\delta ^{i} _\alpha \delta^j _\beta - \delta^i _\beta \delta^j _\alpha ) , 
\ee
where the gap field, $\Delta$, is a complex number which is the order parameter for the condensate phase. The $a,b$ indices are Dirac indices, and we have included the redundant $L$ subscript to make explicit that both quarks are left-handed.

The symmetries of the gap will dictate the structure of the low energy effectively field theory. These are as follows: the gap is a singlet (i.e. antisymmetric) under both the flavour and color SU(2)'s, and hence does not break either symmetry. The gap \emph{is} charged under baryon number (with a charge equal to that of two quarks), and thus the VEV \eqref{eq:psiLpsiL} spontaneously breaks baryon number symmetry $U(1)_B$. It follows that the SU(2) condensate is a \emph{superfluid} \cite{Kogut:2004su, Kogut:1999iv} \footnote{Note however that in both SU(2) QCD and the standard model, baryon number is only an approximate symmetry, and the Goldstone boson of superfluidity will have a small mass. We neglect this mass for the purposes of the present work. This will be explored further in upcoming work \cite{followup}.}. Putting all this together, the symmetry breaking pattern is given by ${\rm SU}(2)_{\rm color} \times {\rm SU}(2)_{\rm flavor} \times {\rm U}(1)_B \rightarrow {\rm SU}(2)_{\rm color} \times {\rm SU}(2)_{\rm flavour}$.

This is slightly different from the scenario of QCD with $N_c = N_f = 3$. In that case, the symmetry breaking pattern is ${\rm SU}(3)_{\rm color} \times {\rm SU}(3)_L\times {\rm SU}(3)_R \times {\rm U}(1)_B \rightarrow {\rm SU}(3)_{\rm color-flavor}$. Both color and flavor symmetry are broken to the diagonal subgroup, and this necessarily also breaks the chiral symmetry \cite{Alford:1998mk} (in contrast with the SU(2) case, where these symmetries are each unbroken\footnote{Note also that the condensate does not generate a mass for the SU(2) quarks. There will, however, be a mass associated with quasiparticle excitations of the condensate.}). The breaking of color symmetry follows from the fact that the SU(3) color anti-symmetric diquark state is the $\overline{\rm 3}$, and hence the condensate is charged under ${\rm SU}(3)$ and is a super\emph{conductor}.

The gap can be computed at weak coupling, and the QCD result is given by \cite{Alford:2007xm},
\be
\label{eq:gap}
\Delta \simeq 10^5 \; \mu \; \frac{1}{g^5} \exp \left( - 3 \pi^2/(\sqrt{2} g) \right),
\ee
where $g$ is the gauge theory coupling.  The scaling with $\exp(-1/g)$, in contrast with the $\exp(-1/g^2)$ scaling in BCS theory, is characteristic of non-Abelian gauge theories and arises due to the long-range exchange of magnetic gluons. For our purposes, we will use equation \eqref{eq:gap} as an estimate of the gap in two-color QCD, and treat the gauge coupling to be a free parameter.

The numerical prefactor in \eqref{eq:gap} is precisely determined and is given in \cite{Alford:2007xm}.  We review the derivation and solution of the gap equation in appendix \ref{gapeqreview}.

\subsection{Fluctuations of the Condensate as Cold Dark Matter}

The relevant low-energy degrees of freedom in the superfluid phase are the Goldstone boson of $U(1)_B$ breaking and the Higgs, which corresponds to fluctuations in the gap itself. The effective field theory can be formulated by integrating out gluons and quarks that are far outside the Fermi sphere, leading to what is known as the `High Density Effective Theory'. The application of this to color-flavor locking was done by Son \cite{Son:2002zn}, and has subsequently been worked out in detail by various authors (see e.g. \cite{Alford:2007xm} and references therein). The inclusion of Higgs-mode fluctuations was incorporated by \cite{Anglani:2011cw}. We provide an overview of this in Appendix \ref{app:flucts}.

 As a complex field, the gap can be expanded as\footnote{Note that $\phi$ appearing here is the Goldstone boson of $U(1)_B$ breaking and not the inflaton.}
\be
\Delta(x,t) = \left[ \Delta_0 + \rho(x,t) \right] e^{2 i \phi(x,t)} 
\ee
where $\Delta_0$ on the right-hand-side is given by the mean-field result, equation \eqref{eq:gap} (we will drop the subscript $0$ from hereon).

Here, $\rho(x)$ is the dark Higgs-mode and $\phi(x)$ is the Goldstone mode of the broken $U(1)_B$ symmetry.  The effective Lagrangian for the dark Higgs-mode $\rho$ is given by \cite{Anglani:2011cw},
\bea
\label{rhoeffaction}
\mathcal{L}_{\rho} = && \frac{1}{2} \frac{3 \mu_{eff}^2}{4 \pi^2} \frac{1}{ \Delta ^2}\left[ (\partial_0 \rho)^2  - \frac{1}{3} (\partial_i \rho)^2\right] \nonumber\\
&&- \frac{1}{2}\frac{12 \mu_{eff}^2}{\pi^2} \rho^2 +\displaystyle \sum_{n\geq2} c_n \frac{\mu_{eff}^2}{\Delta^{n-2}} \rho^n ,
\eea
where $\mu_{eff}$ is the effective chemical potential $\mu_{eff} \equiv \mu - \partial_0 \phi$. The $c_n$ are a set of coefficients parametrizing the higher-dimension operators. 

Baryon number violation is manifest in the time-evolution of $\phi$. The dynamics are simplest at small $\Delta$, and at zeroth order in $\rho$, where the action for $\phi$ is given by \cite{Anglani:2011cw},
\be
\mathcal{L}_{\phi} = \frac{3}{4 \pi^2} \left[ (\mu - \partial_0 \phi)^2 - (\partial_i \phi)^2 \right].
\ee
The equation of motion for $\phi$, ignoring spatial gradients, then enforces $(\mu - \partial_0 \phi) = c$ for a constant $c$. Interestingly, this is not decoupled from the Higgs-mode fluctuations, but is sourced by them via the term $\mu_{eff} ^2 \rho^2$ appearing in eq. \eqref{rhoeffaction}. This leads to a time-dependence in $\partial_0 \phi$, of the form  $\partial_0 (\partial_0 \phi)/\mu_{eff} \simeq \partial_0 \rho / \rho$.

At constant $\phi$ velocity,  the field $\rho$ can be made canonically normalized via the rescaling $\varphi = (\sqrt{3}/2 \pi)({\mu}/\Delta) \rho$, and the effective theory for $\varphi$ is given by:
\be
\mathcal{L} = \frac{1}{2} (\partial \varphi)^2 - V(\varphi),
\ee
with
\be
\label{Vphi}
V(\varphi) = 16 \Delta^2 \varphi^2 + \displaystyle \sum_{n\geq2} \tilde{c}_n  \frac{\Delta^2}{\mu^{n-2}} \varphi^n .
\ee

The second term on the right of \eqref{Vphi} constitutes a set of self-interactions of $\varphi$, as well as interactions with the goldstone boson (via the replacement $\mu \rightarrow \mu_{eff} = \mu - \partial_0 \phi$). For the present purposes we are interested in the mass of $\varphi$ which is given by
\be
m  = 4 \Delta ,
\ee
and hence the mass of Higgs-mode fluctuations are given by the gap. For a chemical potential in the keV range, $\mu \simeq 10 \mbox{eV}$, and $g\simeq 0.3$, this gives,
\be
\label{eq:examplemass}
m = 10^{-22} \mbox{eV} ,
\ee
which is in the mass range of an ultra-light scalar dark matter candidate \cite{Hu:2000ke, Hui:2016ltb}. The requisite phenomenology, e.g. the relic abundance, then directly follows that of the axion literature (see e.g. \cite{Marsh:2015xka}).

We now go over some of the salient phenomenological details of this model. 

\subsection{Phenomenology}

\subsubsection{Dark Matter Self-Interactions and Mass Range }

Dark matter self-interactions are strongly constrained by astrophysics, notably galaxy morphology. The constraints arise because self-interactions isotropize dark matter halos, as discussed in the original work on Self Interacting Dark Matter \cite{Spergel:1999mh}. This is in tension with observational evidence for triaxial structure of dark matter halos \cite{triaxial}. 

The dark matter self-interaction cross-section is constrained to be $\sigma/m \lesssim 0.1 \mbox{g}/\mbox{cm}^2$ \cite{Tulin:2017ara},  where $m$ is the mass of the dark matter particle. For $\lambda \phi^4$ theory, this can be written in terms of the mass and coupling as $ \lambda^2 \left( \frac{8 \mbox{MeV}}{m}\right)^3 < 0.1 $. In our scenario, keeping only the leading order correction to the action, the effective theory has the form
\be
V = \frac{1}{2} m^2 \varphi^2 + \frac{1}{4!}\lambda \varphi^4 ,
\ee
with 
\be
\lambda \simeq \frac{\Delta^2}{\mu^2}. 
\ee
The self-interaction constraint can be written as
\be
\label{eq:selfint}
\frac{m}{\mbox{eV}} < 10^{- 31} \left( \frac{\mu}{\mbox{eV}} \right)^4 .
\ee
The mass \eqref{eq:examplemass} is well within this range, and hence is safe from self-interaction constraints.

More generally, the dark matter mass in this scenario is determined by the chemical potential and the gauge coupling. The gauge coupling is effectively a free parameter, while the chemical potential, produced by primordial gravitational waves, depends on the number density via the relation $\mu = (2 \pi^2 n)^{1/3}$. The result  \eqref{nDs} leads to
\be
\mu \simeq 10^{-4} T_{re} ,
\ee
where $T_{re}$ is the temperature of the universe at reheating. This is bounded from both above and below, since reheating should occur after inflation but before big bang nucleosynthesis. At the high end, instantaneous reheating in high scale inflation (e.g. $m^2 \phi^2$) yields $T_{re} \simeq 10^{16} \mbox{GeV}$, while at the low end, consistency with big bang nucleosynthesis requires $T_{re} \gtrsim \mathcal{O}(1)$ MeV \cite{deSalas:2015glj}. This implies a range of allowed chemical potentials $ 10^2  \mbox{eV} < \mu < 10^{12}\mbox{GeV}$.

The allowed mass range can then be determined from the cosmological history (via the reheat temperature $T_{re}$) and consistency with the self-interaction constraint \eqref{eq:selfint}, i.e.
\be
\frac{m}{\mbox{eV}} < 10^{- 47} \left( \frac{T_{re}}{\mbox{eV}} \right)^4 .
\ee
The scenario of high scale reheating leads to the maximal upper bound on the mass, which is however  above the Planck scale, $m < 10^{53} $ eV.  Thus a wide range of masses are possible: For example, in addition to the ultra-light mass \eqref{eq:examplemass},  an $\mathcal{O}(\mbox{GeV})$ mass arises for $\mu\simeq10$ GeV. 

\subsubsection{Isocurvature Constraints}

In axion dark matter scenarios, constraints can arise from the possibility that the axion becomes dynamical during inflation \cite{Marsh:2015xka}.  These apply for models where the PQ symmetry is broken during inflation: As a (nearly)-massless field, the axion will have vacuum fluctuations that contribute to the CMB spectra.  However, if the PQ symmetry is unbroken during inflation, then the axion is not a dynamical field during inflation and it does not acquire a perturbation spectrum. It follows that in this case the axion does not contribute to the CMB.

Our scenario is analogous to the latter case, since the condensate does not exist until after inflation. Hence there are no isocurvature constraints on DBS dark matter.

There is however another source of isocurvature perturbations in this scenario: the production of gauge fields during inflation.  As studied by one of the authors in \cite{McDonough:2016xvu} (albeit in the Abelian case), this leads to an isocurvature (entropy) perturbation, which can in principle be in conflict with CMB observations, either via direct isocurvature constraints \cite{Abazajian:2016yjj} or else indirectly via the amplitude of the primordial power spectrum \cite{Moghaddam:2014ksa}. 

In the hydrodynamical formulation of cosmological perturbation theory, the entropy perturbation can be described via the resulting non-adiabatic pressure perturbation:
\begin{equation}
\delta P_{nad} \, = \, \dot{p_\phi} \left( \frac{\delta p_A}{\dot{p}_\phi} - \frac{\delta \rho_A}{\dot{\rho}_\phi} \right) \, ,
\end{equation}
where $\delta p_A$ and $\delta \rho_A$ are the pressure and energy density fluctuations in the gauge field. This sources the curvature perturbation $\zeta$ via the relation \cite{Malik:2004tf}
\begin{equation}
\dot{\zeta} \, = \, - \frac{H}{p+\rho} \delta P_{nad}  .
\end{equation}
It was found in \cite{McDonough:2016xvu} that this leads, during preheating, to an efficient conversion of the entropy perturbation into curvature perturbations, obviating direct isocurvature constraints, and the resulting curvature perturbation is well within observational constraints.

\section{Discussion}\label{SEC.Discussion}

In this work we have presented a method for achieving a particle-anti-particle asymmetry in dark and visible matter. Contrary to existing scenarios, the dark matter asymmetry is produced \emph{primordially}, occurring via the axial-gravitational anomaly in the presence of chiral gravitational waves. As a result, the mechanism here requires no annihilation thermal freeze-out nor transfer of asymmetry from the standard model sector to the dark sector, and hence requires no couplings to the standard model. 

The simplest model realizing this mechanism leads to a qualitatively new class of dark matter models: dark baryon-charge superfluid (DBS) dark matter. While superficially resembling the Bose-Einstein Condensate and superfluid dark matter scenarios, the DBS models, via their realization in non-abelian gauge theory, yield a rich structure not present in existing constructions. 

The effective potential of our model, $V = m^2 \varphi^2 + \lambda \varphi^4$, is the same used by axion dark matter (see \cite{Marsh:2015xka} for a review), and in axion Bose-Einstein Condensate Dark Matter \cite{Sikivie:2009qn}. Similar to vanilla axion DM \cite{Marsh:2015xka}, the dark matter in our scenario is never in thermal equilibrium with the rest of the universe, making it a viable cold dark matter candidate.

However, despite some overlap in terminology, our setup is fundamentally different from existing works: In our model, $\varphi$ is  itself an excitation of a dark baryon charge condensate. This is contrast with axion BEC DM \cite{Sikivie:2009qn}, where it is excitations of $\varphi$ that condense to form the condensate. In our scenario, excitations need not form a BEC or rejoin the condensate. Moreover, while dark matter in our scenario is a superfluid, it is fundamentally different from the superfluid dark matter of Khoury and Berezhiani \cite{Berezhiani:2015bqa}, where strong self-interactions lead to a superfluid phase inside of halos; in our scenario the self-interactions are weak. In addition, our scenario includes the Goldstone boson of $U(1)_B$ breaking, which has intricate couplings to $\varphi$. The dynamics of this Goldstone boson may yield its own distinct signatures.

The greatest advantage of the scenario presented here over existing models is its UV completion in non-Abelian gauge theory. This opens up a wealth possibilities for extensions to the model: 
 \begin{itemize}
\item[1.] Additional flavours of massive quarks, which do not condense (similar to the strange quark in the ``2SC'' phase of QCD \cite{Alford:2007xm}), but rather lead to a hadronic component of dark matter
\item[2.] The presence of color-charged vortices \cite{vortices} inside of dark matter halos 
\item[3.] Direct couplings of the dark quarks to visible quarks, leading to semi-visible jets at the LHC \cite{Cohen:2015toa}
\end{itemize}
None of these extensions arise in axion models. Thus the dark baryon superfluid scenario presented here opens up a whole new phenomenology.

An interesting avenue for further research is the realization of the Baryonic Tully-Fisher relation \cite{Tully:1977fu} via irrelevant operators that couple the dark quarks to visible sector quarks. In this way, DBS could provide a first principles derivation of the condensate-baryon interaction $\varphi \rho_b$ utilized for this purpose in \cite{Berezhiani:2015bqa}. We intend to explore this possibility, and others (e.g. a covariant derivation of the gap equation \cite{Alexander:2009uu} and the structure of DM halos \cite{Alexander:2016glq}), in future work.  Finally, the low energy dynamics of the condensate involves non-trivial derivative interactions between the Dark-Higgs and Nambu-Goldstone mode which will require numerical treatment.  We will pursue a study of the formation and morphology of halos in the presence of these new interactions in the future.

\section*{Acknowledgements}

The authors thank Robert Brandenberger, Elisa Ferriera, Jiji Fan, Jim Gates, Justin Khoury, David Pinner, Robert Sims, Devin Walker and Neal Weiner for useful discussions. EM is supported in part by a Post-Doctoral Fellowship from the National Science and Engineering Research Council of Canada.

\appendix
\section{Gravitational Waves and Leptogenesis in Chromo-Natural Inflation}
\label{app:GWs}
In order to compute the dark quark number density produced after inflation, we need to evaluate the gravitational wave mode functions, sourced by vacuum fluctuations in the metric and gauge field. The metric fluctuations correspond to transverse traceless fluctuations to the spatial metric of spacetime. For a gravitational wave propogating in the $z$-direction, the metric takes the form
\bea
\mathrm{d}s^2 = - &&\mathrm{d}t^2 + a^2(t)\left[\right. (1 - h_+) \mathrm{d}x^2 \nonumber \\
&& + (1 + h_+)\mathrm{d}y^2 + 2 h_\times \mathrm{d}x \mathrm{d}y + \mathrm{d}z^2 \left. \right],
\eea
where $a(t)$ is the scale factor, and $h_{+/\times}$ are the two polarizations of the graviton. The gravitational waves can be wirtten in a circular polarization basis via the rotation
\be
\label{eq:hLR}
h_{L} = (h_{ +} - i h_\times)/\sqrt{2} \;\; ,\;\; h_{R}=(h_+ + i h_\times)/\sqrt{2} .
\ee
It will  be useful work with a re-scaled version of \eqref{eq:hLR}, $v_{L,R}$ defined via 
\be
h_L =  \frac{v_L}{M_{pl} a} \;\;,\;\; h_R =  \frac{v_R}{M_{pl} a} .
\ee

There are also tensorial fluctuations in the gauge field sector, $t_{+/-}$, 
\begin{equation}
\delta A^a_\mu = a\, t^a_\mu = a \left( \begin{array}{cccc}
0 & t_+ & t_\times & 0 \cr
0 & t_\times & -t_+ & 0 \cr
0 & 0 & 0 & 0
\end{array}\right) ,
\label{eqn:Atnsr}
\end{equation}
which again we rotate and rescale as
\be
t_L =  \frac{u_L}{M_{pl} a} \;\;,\;\; t_R =  \frac{u_R}{M_{pl} a} .
\ee
The power spectrum of gravitational waves is given by that sourced by both $u$ and $v$. The result is simply, 
\bea
&&\langle h_{L} h_{L} \rangle = (2 \pi)^3 \delta(k+k') P_{L}(k) \\
&& P_{L}(k) = \frac{4}{a^2 M^2 _{Pl}}(|v_L|^2 + |u_L|^2)
\eea
and similarly for $R$.

The equations of motion for $v_{L,R}$ and $u_{L,R}$ can be straightforwardly worked out:
\begin{eqnarray}
&&v_L'' + \left[k^2 - \frac{a''}{a}  + \frac{2}{a^2 M_P^2}(g_A ^2 \sigma^4 - \sigma'^2)\right] v_L = \nonumber \\
&&\frac{2}{a M_P}\left[ (g_A \sigma + k) g_A  \sigma^2 u_L -  \sigma' u_L'\right] ,
\label{eqn:vL} \\
&&u_L'' + \left[k^2 + 2 g_A k \sigma +4(g_A \sigma + k)\frac{\phi'}{M}\right]u_L  =\nonumber
\\
&&\frac{2}{a M_P}\left[ a\left(\frac{v_L}{a}\right)' \sigma' + g_A \sigma^2 \left(k-g_A\sigma +4 \frac{\phi'}{M} \right)  v_L  \right].
\label{eqn:uL}
\end{eqnarray}
where $'$ indicates a derivative with respect to conformal time. 

The gauge field fluctuations have a tachyonic instability, for only one chirality, for $k$ satisfying
\be
k^2 + 2 g_A k \sigma + 4 (g_A \sigma + k) \frac{\phi'}{M} < 0 ,
\ee
This tachyonic growth leads to large-scale gravitational waves sourced by the gauge field. On the other hand, the metric tensor fluctutations have no instability, but instead receive only an additional (positive) mass. Hence the dominant source of gravitational waves is the gauge field fluctuations.

Assuming that inflation begins with a negligible asymmetry, the standard model leptogenesis result is given by  \cite{Caldwell:2017chz}
\bea
\label{eq:result}
&& \langle n_{\ell} \rangle =  \frac{1}{64 \pi^2 a^3} \int \mathrm{d}\log k \left[ k^3 (\Delta_R^2 - \Delta_L ^2) - k ({ \Delta_R '} ^2 - {\Delta_L '} ^2)\right]  \nonumber \\
&&\Delta_P ^2 = \frac{k^3}{\pi^2} \left( |h_{P,k}|^2 + |h_{P,k}|^2 _{g}\right) \nonumber \\
&&{ \Delta_P'} ^2 = \frac{k^3}{\pi^2} \left( {|h'_{P,k}|}^2 + |{h'_{P,k}}|^2 _g \right) ,
\eea
where $n_\ell$ is the number density of standard model leptons, $P=L,R$ is the handedness, and the subscript $g$ refers to tensorial fluctuations of the gauge field $F$. 

This can be evaluated and expressed as a ratio with the entropy density of the universe, $s = 2 g_{*s}\pi^2 T^3/45$, as
\be
\frac{|\langle n_{\ell} \rangle|}{s} = 2.45 \times 10^{-10} .
\ee
where the parameters chosen are $g_{*s}= 1.4 \times 10^{-3}$, $M=1.7 \times 10^{-4} M_{Pl}$, and the inflation model is given by and the fiducial inflation model $V =m^{4}(\phi/m)^n /n$ with $m = 1.7 \times 10^{-3}M_{Pl}$ and $n=1/8$.

\section{Review of the Gap Equation}
\label{gapeqreview}

A detailed derivation of the gap equation for QCD at finite chemical potential can be found in the review \cite{Alford:2007xm} and references therein. Here we will review the key aspects of  \cite{Alford:2007xm}, highlighting the differences between QCD and BCS theory.

The energy gap $\Delta$ is a shift in the dispersion relation of excitations, $E_k^2 = \epsilon_k ^2+ \Delta_k ^2$. This manifests itself in field theory as the anomalous self energy of the fermion propagator, induced by interactions with gluons. Schematically the gap appears in the propagator as \cite{Kogut:2004su}
\be
\langle \psi^\dagger _a (p) \psi _a(p)\rangle = \frac{- i p_0 + \epsilon_p}{p_0^2 + \epsilon_p ^2 + \Delta^2_p}
\ee
where $\epsilon_p = |\vec{p}| - \mu$ and $a$ is the flavor index. This can be rearranged to compute $\Delta$ in terms of loop contributions to the propagator, leading to an expression for $\Delta$ that is an integral over quark-gluon interaction vertices and gluon propagators. For single gluon exchange, this takes the form
\be
\Delta \simeq g^2 \int  \mathrm{d}^4q \; v_\mu (q)  D_{\mu \nu}(q-k) v_\nu (-q)
\ee
where $v_\mu (q)$ is the (dressed) quark-gluon vertex and $D_{\mu \nu}$ is the gluon propagator. This is referred to as the ``gap equation.''

The gap equation for QCD can be rigorously computed by deriving the two-particle irreducible (2PI) effective action $\Gamma$.  The self-energy (i.e. the gap) is given by the variation of $\Gamma$ with respect to the fermion propagator. In contrast with BCS theory, in QCD the attractive interaction responsible for the pairing (single gluon exchange between quarks) is present at weak coupling, and the gap equation can be derived using perturbative quantum field theory. 

The gap equation for single gluon exchange, in the simple scenario where there is only one gap parameter, is given by \cite{Alford:2007xm}
\be
\label{gapeq}
\Delta_{k} = \frac{g^2}{4} \int \frac{d^3 q}{(2 \pi^3)} \, Z(q) \frac{\Delta_{q}}{\epsilon_{q}} \left[ D_e(p) T_e + D_m (p)T_m \right] 
\ee
where $Z$ is the wavefunction renormalization, $D_{e,m}$ are the electric and magnetic gluon propagators, and the factors  of $T_{e,m}$ come from traces over color, flavor, and Dirac indices. 

The dominant contribution to the integral comes from gluons with $q_0 \ll q$, i.e. nearly static gluons. The propagators in this limit are given in Coloumb gauge by \cite{Alford:2007xm, Kogut:2004su}
\bea
D_{e}(q) &=& \frac{2}{|\vec{q}|^2 + m_{e}^2} \\
D_{m}(q) &=& \frac{1}{|\vec{q}|^2 +  (3 \pi/4)m_e^2 (q_0/|\vec{q}|)}
\eea
where $m_e$ is the Debye mass, $m_e \simeq g^2 \mu^2$, which screens electric gluons. The magnetic gluons, however, are not screened at all for nearly static fields; the magnetic gluons are \emph{damped} rather then screened, and this occurs on a characteristic scale $|\vec{q}| \sim (g^2 \mu^2 \Delta)^{1/3}$.

Focusing on the magnetic gluon contribution, the gap equation is of the schematic form
\be
\Delta \propto g^2 \int \mathrm{d}\xi \mathrm{d}\theta \; \frac{\Delta}{\sqrt{\xi^2 + \Delta^2}} \cdot \frac{\mu^2}{\theta \mu^2 + \delta^2}
\ee
where $\xi\equiv k-\mu$, $\theta$ is the angle between the loop and external momenta, and $\delta \simeq (g^2 \mu^2 \Delta)^{1/3}$ is the cutoff due to Landau damping. The non-trivial angular integration is particular to QCD, and is not present in BCS theory. 

Performing the angular integration, the gap equation \eqref{gapeq} is given
\be
\Delta_k = \frac{g^2}{18 \pi^2} \int \mathrm{d}q	 \; \frac{\Delta_q}{\epsilon_q} \frac{1}{2} \; \log \left( \frac{\mu^2}{|\epsilon_q ^2 - \epsilon_k ^2|}\right)
\ee
The solution to this is given by
\be
\Delta \propto g^{-5} \exp \left( - \frac{3 \pi^2}{\sqrt{2}g}\right) ,
\ee
as per equation \eqref{eq:gap}. This differs from BCS theory in the power of $g$ appearing in the exponent; in BCS theory the gap scales as $\exp(-1/g^2)$.

\section{Effective Action for Fluctuations of the Gap}

\label{app:flucts}

The previous appendix derived the gap by evaluating loop diagrams in QCD and computing the gluon-exchange induced self energy in the quark propagator. The resulting gap is a constant, and is the mean-field value. To study \emph{fluctuations} in the condensate requires the formulation of a quantum effective action of the low energy degrees of freedom, namely the Goldstone boson of $U(1)_B$ breaking and the fluctuations of the magnitude of the gap. 

The effective action for fluctuations of the gap can be derived using a combination of the Nambu-Jones-Lasinio model, where the quark-gluon interaction is modeled as a four-fermi interaction, and high density effective theory (HDET). Here we will overview the analysis of \cite{Anglani:2011cw}.

In this procedure, one allows the gap to fluctuate about its mean-field value $\overline{\Delta}$, and computes the coupling of fermions to the quasiparticle fluctuations. One then integrates out the fermions and arrives at the effective action for the fluctuations.

In general, the gap is a matrix in color-flavor space and can be expanded as
\begin{equation}
\Delta_{AB}(x) \rightarrow [\Delta_{AB}+\rho_{AB}(x)]e^{2i\phi(x)}\,,
\end{equation}
where the indices $A,B$ are indices in color-flavor space, and $\Delta$ on the right-hand-side is the mean-field value. Note that $\phi$ appearing here is the Goldstone boson of $U(1)_B$ breaking and not the inflaton.

The expanded mean-field Lagrangian is then given by,
\bea\label{expansion}
{\cal L} &=& \bar \psi \left(i \gamma^\mu \partial_{\mu}+
\gamma^0\mu -\gamma^{0}\partial_{0}\phi-\gamma^i \partial_i  \phi \right) \psi \nonumber\\ &-&\frac{1}{2} \psi^{\dagger}
C(\Delta +
\rho)\psi^* +\frac{1}{2} \psi^{t} C(\Delta + \rho)  \psi\,. 
\eea
After integrating out the fermions, the end result is
\bea
{\cal S}_{\rm eff}&=&-\frac{i}{g}\int
d^{4}x\left[\rho_{AB}(x)\rho^{AB}(x)+2\rho_{AB}(x)\Delta^{AB}\right] \nonumber\\ &-&\frac{i}{2}{\rm
Tr}\ln\left(1+S_{MF}\Gamma\right)\,,
\eea
where the $AB$ indices are contracted with a tensor $W_{ABCD}$ determined by the index structure of the diquark condensate, and the second line contains as expansion 
\be
{\rm Tr}\ln\left(1+S_{MF}\Gamma\right)={\rm
Tr}\left[\sum_{n=1}^{\infty}\frac{(-1)^{n+1}}{n}(S_{MF}\Gamma)^{n}\right]\,,
\ee
where $\Gamma$ is the interaction vertices of the gap fluctuations with the fermions. 

The expression for $\Gamma$ is the main result of the integrating-out procedure, and is schematically given by $\Gamma = \Gamma_{\rho} + \Gamma_1 ^\mu A_\mu + \Gamma_2 ^{\mu \nu} A_{\mu \nu} + \Gamma_3 ^{\mu \nu \sigma} A_{\mu} A_{\nu} A_{\sigma} + \Gamma_4 ^{\mu \nu \sigma \delta} A_{\mu} A_{\nu} A_{\sigma} A_{\delta}$, where $\Gamma_{i}$ are coefficients given in \cite{Anglani:2011cw}, and the gauge field $A_{\mu}$ is a repackaging of the Goldstone boson: $A^\mu = (\partial_0 \phi, \nabla \phi)$.

As an example of the utility of this, consider the piece of the effective action proportional to $\partial_0 \phi$. This corresponds to 
\be
{\cal L}_1 = -i  {\rm Tr}[S \Gamma]\Big|_{1} \,.
\ee
where the subscript $1$ indicates that only linear terms in the gauge field are kept. The resulting expression is given by
\be
{\cal L}_1 = \left(-\frac{3}{\pi^2}\mu^3+\frac{6}{\pi^2}\mu^2\Delta\right) \partial_0 \phi  \,.
\ee
The same method can be applied to the Higgs-mode fluctuation $\rho$, which leads to the effective action \eqref{rhoeffaction}.

\end{document}